\newcommand{\eps}{\epsilon}

\documentclass[prl, amssymb, twocolumn, showkeys, showpacs]{revtex4}
\usepackage{graphicx}
\begin{document}

\title{Equilibrium Distribution of Mutators in the Single Fitness Peak Model}

\author{Emmanuel Tannenbaum}
\email{etannenb@fas.harvard.edu}
\author{Eric Deeds}
\author{Eugene I. Shakhnovich}
\affiliation{Harvard University, Cambridge, MA 02138}

\begin{abstract}

This paper develops an analytically tractable model for determining 
the equilibrium distribution of mismatch repair deficient strains in 
unicellular populations.  The approach is based on the single fitness 
peak (SFP) model, which has been used in Eigen's quasispecies equations 
in order to understand various aspects of evolutionary dynamics.  As with 
the quasispecies model, our model for mutator-nonmutator equilibrium
undergoes a phase transition in the limit of infinite sequence length.  
This ``repair catastrophe'' occurs at a critical repair error probability of 
$ \eps_r = L_{via}/L $, where $ L_{via} $ denotes the length of the genome 
controlling viability, while $ L $ denotes the overall length of the genome.
The repair catastrophe therefore occurs when the repair error probability 
exceeds the fraction of deleterious mutations.  Our model also gives a 
quantitative estimate for the equilibrium fraction of mutators in 
{\it Escherichia coli}.  

\end{abstract}


\maketitle

In order to preserve the integrity of their genomes, living systems have
evolved sophisticated mechanisms for correcting errors in their DNA sequences
\cite{VOET}.  Otherwise, genetic damage due to environmental factors such as 
radiation, metabolic free radicals, and mutagens, combined with replication 
errors, would lead to unviable organisms due to the unrecoverable loss of 
genetic information.  This phenomenon, which was first characterized by Eigen 
in \cite{EIG1}, is known as the ``error catastrophe'' \cite{EIG1, EIG2}.  It 
has since been studied in a number of theoretical papers \cite{SFP1, SFP2,
QUAS1} (and references therein), and has also been observed 
experimentally \cite{CAT1}. 

Some of the error-correcting ability in living systems is already built into 
the DNA polymerases themselves.  In {\it Escherichia coli}, the proofreading 
ability of the DNA polymerases Pol I and Pol II results in an error 
probability of $ 10^{-6} - 10^{-7} $ per base pair \cite{VOET}.  Additional 
enzymes continuously scan the DNA molecule, repairing lesions and 
mismatches that occur due to environmental damage.  

A key error-repair mechanism is known as mismatch repair, and occurs 
immediately following DNA replication.  The mismatch repair system scans 
the DNA molecule, identifies, and then corrects mismatched base pairs.
Mismatch repair in {\it E. coli} reduces the error probability in DNA 
replication to $ 10^{-8} - 10^{-10} $ per base pair \cite{VOET}.  Cells with 
inactivated mismatch repair consequently have mutation rates which are 
$ 10 $ to $ 10,000 $ times higher than cells whose mismatch repair system is 
functioning.  Because of their higher than wild-type mutation rates, these
``mutator'' strains are believed to play an important role in the emergence of 
antibiotic resistance, and cancer in multicellular organisms 
\cite{MUT1, MUT2, MUT3, MUT4, MUT5, MUT6}.

To develop a model for the equilibrium distribution of mutators versus
nonmutators in a unicellular population, we consider a genome of
alphabet size $ S $ (``bases'' $ 0, 1, \dots, S-1 $), consisting of 
two genes.  The first gene consists of $ L_{via} $ bases, and controls the 
viability of the genome.  The second gene consists of $ L_{rep} $ bases, and 
codes for the enzymatic machinery involved in mismatch repair.  If we let 
$ \sigma $ denote an arbitrary gene sequence, then we may write, 
$ \sigma = \sigma_{via} \sigma_{rep} $.

We assume a single fitness peak (SFP) model for both genes.  
Thus, there is a unique, ``fit'' sequence $ \sigma_0 = \sigma_{via, 0}
\sigma_{rep, 0} $.  A cell with genome $ \sigma $ has a first-order growth 
rate constant $ k >> 1 $ if $ \sigma_{via} = \sigma_{via,0} $,
and $ 1 $ otherwise.  Mismatch repair has an error probability of 
$ \eps_r $ per mismatched base pair, and is functioning only when
$ \sigma_{rep} = \sigma_{rep,0} $.

While somewhat artificial, the SFP model has been successfully applied in 
\cite{SFP3} toward understanding the correlations between antibody and viral
mutation rates.  Furthermore, because proteins generally have a key set of
conserved residues which more or less dictate their final structure and 
function, the corresponding gene has a subsequence of conserved bases required
for its proper function \cite{MIRNY1, VOET}.  Thus, by summing over the 
unconserved bases, it is possible to reduce the fitness landscape to an SFP in 
the conserved subsequence.  Therefore, there is reason to believe that many of 
the phenomenological aspects of our system can be captured by an SFP-based 
approach, and that such an approach can also be semiquantitative in a number 
of cases.

The basic equation governing the dynamics on the genome space has the form
of Eigen's quasispecies equations \cite{EIG1, EIG2},
\begin{equation}
\frac{dx_{\sigma}}{dt} = (\kappa_{\sigma} - \bar{\kappa}(t)) x_{\sigma} +
\sum_{\sigma' \neq \sigma}{[\kappa_m(\sigma', \sigma) x_{\sigma'} -
\kappa_m(\sigma, \sigma') x_{\sigma}]}
\end{equation} 
where $ x_{\sigma} $ denotes the fraction of the population with genome
$ \sigma $, $ \kappa_{\sigma} $ is the growth rate constant of $ \sigma $,
$ \kappa_m(\sigma, \sigma') $ denotes the mutation rate constant from 
$ \sigma $ to $ \sigma' $, and $ \bar{\kappa}(t) \equiv 
\sum_{\sigma}{\kappa_{\sigma} x_{\sigma}(t)} $.

We assume that replication errors are sufficiently small so that we 
need only worry about point mutations, and that since $ k >> 1 $, any flow
off of the viability peak is unidirectional.  Furthermore, since we are only
interested in the relative distribution of viable mutators and 
non-mutators, we only focus on a subspace of sequences given by 
$ \sigma = \sigma_{via,0}\sigma_{rep} $.  Thus, from now on, we shall 
simplify matters and redenote $ \sigma_{rep} $ as $ \sigma $.  The full genome
is $ \sigma_{via, 0}\sigma $ by implication.  On this subspace, the effective 
growth rate constant becomes $ k(1 - L_{via}\eps_{\sigma}) $, due to
leakage off of the fitness peak.  Here $ \eps_{\sigma} $ denotes the
per base pair replication error probability, and is equal to $ \eps 
\eps_r $ if $ \sigma = \sigma_0 $, and $ \eps $ otherwise.  Finally, by the 
symmetry of our system, we may make the further assumption that $ x_{\sigma} $ 
depends only on the Hamming distance $ HD(\sigma, \sigma_0) $ from 
$ \sigma_0 $.  Thus, defining $ \Omega_l(\sigma) = \{\sigma'| HD(\sigma', 
\sigma) = l\} $, we may then also define $ x_l = x_{\sigma} $, where 
$ \sigma \in \Omega_l(\sigma_0) $.  Note that point mutations between $ x_l $ 
and some $ x_{\sigma} $ may only occur if $ x_{\sigma} \in \Omega_{l, l \pm 1}
(\sigma_0) $.  However, we may negect intra-$ \Omega_{l}(\sigma_0) $ couplings,
due to cancellation of mutational inflows and outflows.  

A $ \sigma \in \Omega_{l}(\sigma_0) $ may be connected via a point mutation to 
a $ \sigma' \in \Omega_{l-1}(\sigma_0) $ by changing any one of the $ l $ 
bases distinct from the corresponding bases in $ \sigma_0 $ back to the 
corresponding base in $ \sigma_0 $.  Thus, there are $ l $ possible 
connections.  A $ \sigma \in \Omega_{l}(\sigma_0) $ may be connected via a 
point mutation to a $ \sigma' \in \Omega_{l+1}(\sigma_0) $ by changing any one 
of the $ L_{rep} - l $ bases equal to the corresponding bases in $ \sigma_0 $.
Since there are $ S - 1 $ possibilities per base, the result is 
$ (L_{rep} - l)(S - 1) $ connections.  The net mutational flow is then,
\begin{widetext}
\begin{equation}
\sum_{\sigma' \neq \sigma}{[\kappa_m(\sigma', \sigma) x_{\sigma'} -
\kappa_m(\sigma, \sigma') x_{\sigma}]} = \frac{kl}{S-1}(\eps_{l-1} x_{l-1} - 
\eps_l x_l) + k(L_{rep} - l)(\eps_{l+1} x_{l+1} - \eps_l x_l)
\end{equation}
\end{widetext}
where $ \eps_0 = \eps \eps_r $, and $ \eps_l = \eps $ for $ l \geq 1 $.  We 
divide the $ \eps $'s by $ S-1 $ because a point mutation can occur to any one
of the $ S-1 $ bases distinct from the changed base.

We also have, $ \kappa(t) = k(1 - L_{via} \eps) + k L_{via} \eps (1 - \eps_r) 
x_0 $.  Now, define $ C_l = {L_{rep} \choose l}(S-1)^l $, the number of 
elements in $ \Omega_l(\sigma_0) $, and set $ z_l = C_l x_l $.  If we reexpress
the dynamical equations in terms of $ z_l $, then at equilibrium we obtain
the system of equations,
\begin{widetext}
\begin{eqnarray}
0 & = & \frac{L_{via}}{L_{rep}}(1 - \eps_r) z_0 (1 - z_0) + 
        \frac{z_1}{L_{rep} (S-1)} - \eps_r z_0 \nonumber \\
0 & = & -\frac{L_{via}}{L_{rep}}(1 - \eps_r) z_0 z_1 +
         \eps_r z_0 - (1 - \frac{1}{L_{rep}} + \frac{1}{L_{rep}(S-1)}) z_1 +
	 \frac{2}{L_{rep}(S-1)} z_2 \nonumber \\
\vdots \nonumber \\
0 & = & -\frac{L_{via}}{L_{rep}}(1 - \eps_r) z_0 z_l +
         (1 + \frac{1}{L_{rep}} - \frac{l}{L_{rep}}) z_{l-1} -
	 (1 - \frac{l}{L_{rep}} + \frac{l}{L_{rep}(S-1)}) z_l +
	 \frac{l+1}{L_{rep}(S-1)} z_{l+1} \nonumber \\
\vdots \nonumber \\
0 & = & -\frac{L_{via}}{L_{rep}}(1 - \eps_r) z_0 z_{L_{rep}} +
        \frac{z_{L_{rep} - 1}}{L_{rep}} - \frac{z_{L_{rep}}}{S-1}
\end{eqnarray} 
\end{widetext}

Except for the last equation, the $ (l+1)^{\mbox{st}} $ equation has a 
mutational contribution from $ z_{l+1} $ which scales as $ 1/L_{rep} $.  
This means that the contribution to $ z_l $ due to backmutation from 
$ z_{l+1} $ becomes negligible for large sequence lengths.  This makes sense, 
since for finite $ l $, the ratio $ C_{l'}/C_l \rightarrow \infty $ as 
$ L_{rep} \rightarrow \infty $ for $ l' > l $, so the probability of mutating 
to lower values of $ l $ vanishes in the limit of infinite sequence length.

The term, $ (L_{via}/L_{rep})(1 - \eps_r) z_0 (1 - z_0) $, and the 
corresponding terms in the other equations arise from the $ \bar{\kappa}(t) $ 
term in the original quasispecies equations (Eq. (1)) of our model.  Because 
the nonmutator sequence has a lower level of leakage off of the viability peak
than the mutator sequences, it has a higher effective growth rate constant.
Thus, when dealing with population fractions as opposed to absolute 
populations, the result is an effective positive flow into the nonmutator
sequence given by the term $ (L_{via}/L_{rep})(1 - \eps_r) z_0 (1 - z_0) $.
By a similar argument, it can be seen why the corresponding terms in the
remaining equations are negative.

We wish to solve these equations for a fixed value of $ \alpha \equiv
L_{via}/L_{rep} $ in the limit of infinite sequence length $ L $.
Let us focus first on the behavior of $ z_0 $ in this limit.  In the first
equilibrium equation, the $ z_1 $ term drops out as $ L_{rep} \rightarrow
\infty $, giving,
\begin{equation}
0 = z_0 (\alpha (1 - \eps_r) (1 - z_0) - \eps_r)
\end{equation}
which has the solutions $ z_0 = 0, 1 - \eps_r/(\alpha (1 - \eps_r)) $.  
The first solution is inconsistent with the requirement that $ z_0 = 1 $
when $ \eps_r = 0 $.  However, the second solution only holds as long as 
$ z_0 \in [0, 1] $.  Clearly, $ z_0 \leq 1 $ $ \forall $ $ \eps_r $.  The 
other requirement that $ z_0 \geq 0 $ gives,
\begin{equation}
\eps_r \leq \frac{\alpha}{1 + \alpha} = \frac{L_{via}}{L}
\end{equation}
Defining $ \eps_{r, crit} = \frac{\alpha}{1 + \alpha} $, we see that in
the limit of infinite genome length, our system has two ``phases.''  For
$ \eps_r < \eps_{r, crit} $ the system is in a ``non-mutator,'' or, 
equivalently, ``repairer'' phase, in which the fraction of non-mutators is a 
quantity which depends only on $ \alpha $ and $ \eps_r $.  At $ \eps_r = 
\eps_{r, crit} $ the system undergoes a ``phase'' transition, which we 
term the ``repair catastrophe,'' after which the system is in a ``mutator'' 
(``non-repairer'') phase.  In this phase, there is essentially no preference 
for being a non-mutator, and the fraction of non-mutators becomes inversely 
proportional to the total number of gene sequences.

A key parameter to study in the phase behavior of our model is the localization
length, given by,
\begin{equation}
\langle l \rangle = \sum_{l = 1}^{L_{rep}}{l z_l}
\end{equation}
This quantity measures the mean Hamming distance of the population from the
nonmutator sequence.  To compute $ \langle l \rangle $ below the phase 
transition in the limit of $ L_{rep} \rightarrow \infty $, we may note that 
for finite $ l $ our equilibrium equations become,
\begin{eqnarray}
0 & = & \alpha (1 - \eps_r) z_0 (1 - z_0) - \eps_r z_0 \nonumber \\
0 & = & -\alpha (1 - \eps_r) z_0 z_1 + \eps_r z_0 - z_1 \nonumber \\
\vdots \nonumber \\
0 & = & -\alpha (1 - \eps_r) z_0 z_l + z_{l-1} - z_l \nonumber \\
\vdots 
\end{eqnarray}
We have already solved the first equation.  The next two equations can be
solved together to give, for $ l \geq 1 $, 
\begin{equation}
z_l = \eps_r (1 + \alpha (1 - \eps_r) z_0)^{-l} z_0
\end{equation}
It should be noted that, while each $ z_l $ converges to the corresponding 
formula given above as $ L_{rep} \rightarrow \infty $, the convergence is not 
uniform, since the larger the $ l $, the larger $ L_{rep} $ must be made to 
get $ z_l $ within some specified cutoff of its $ L_{rep} = \infty $ value.

Define $ z_{l,\infty} = \lim_{L_{rep} \rightarrow \infty}{z_l} $.  It may be 
readily checked that $ \sum_{l = 0}^{\infty}{z_{l, \infty}} = 1 $, so total 
population is conserved in this limiting process.  The localization length 
is given by,
\begin{equation}
\langle l \rangle = \sum_{l = 1}^{\infty}{l z_{l, \infty}} =
\frac{1 - \eps_{r, crit}}{\eps_{r, crit}}\frac{\eps_r}{\eps_{r, crit} - \eps_r}
\end{equation}
Note that, as expected, the localization length is finite for $ \eps_r 
< \eps_{r, crit} $, but diverges at the phase transition.

It is also useful to solve the equilibrium distribution exactly for the case 
$ \alpha = 0 $.  This corresponds to $ L_{rep} = L $, that is, the entire 
genome consists of the repair gene.  Note that $ \eps_{r, crit} = 0 $, so that
the system is always in the mutator phase.  In this case, it may be shown that
the equilibrium solution is given by, $ z_l = {L_{rep} \choose l} (S - 1)^l 
\eps_r z_0 $ for $ l > 0 $, and so the requirement that 
$ \sum_{l = 0}^{L_{rep}}{z_l} = 1 $  gives, $ z_0 = 1/(1 + \eps_r 
(S^{L_{rep}} - 1)) $.  It is readily shown that for large $ L_{rep} $, the 
localization length $ \langle l \rangle \rightarrow (1 - 1/S) L_{rep} $.  This 
result is equivalent to the case of a uniform distribution, which makes sense 
since for $ \alpha = 0 $ there is no preference for being a nonmutator in the 
limit of large $ L_{rep} $.

The phase behavior which emerges from this model may be understood as follows:
For highly efficient repair, the selective advantage for being a nonmutator
is sufficiently large to cause the population to equilibrate in a localized
cluster about the nonmutator sequence.  When repair is inefficient, the
accumulation of deleterious mutations in both mutators and nonmutators is
comparable, and hence the mutators, which are entropically strongly favored,
dominate the population.  The selective advantage for being a nonmutator
is dictated by $ \alpha $, since for low $ \alpha $ there is relatively little
leakage off of the fitness peak, while for high $ \alpha $ there is a large
amount of leakage off of the fitness peak.  Thus, for low $ \alpha $, 
repair has to be highly efficient to give the nonmutators a sufficient 
selective advantage to be in the nonmutator phase, while for high $ \alpha $,
nonmutators have a significant advantage even for relatively inefficient 
repair.  

One of the main features to note regarding the mutator-nonmutator equilibrium 
is that it is independent of the background error probability $ \eps $.
This feature is interesting because as $ \eps \rightarrow 0 $, the difference
in viability between the mutators and the nonmutators disappears.  Thus,
one might naively expect $ \eps_{r, crit} $ to be a function of $ \eps $,
but in the limit of small $ \eps $ (so that only point mutations are 
important), this is not the case.

Figure 1 shows a plot of $ z_0 $ versus $ \eps_r $ for $ \alpha = 1/9, 1, $
and $ 9 $.  We used $ S = 2 $, and took a value of $ L_{rep} = 1,000 $
in order to sufficiently converge the calculations.  The equilibrium equations 
were solved by using fixed-point iteration at every $ \eps_r $.  Note that the 
phase transition does indeed occur at the predicted values of 
$ \eps_{r, crit} $.  The analytical, $ L_{rep} = \infty $ curves lie 
essentially on top of our numerical results, and were therefore not plotted 
here.

\begin{figure}
\includegraphics[width = 0.6\linewidth, angle = -90]{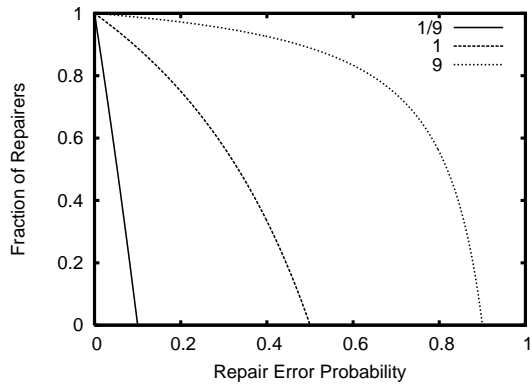}
\caption{Plots of $ z_0 $ versus $ \eps_r $ for $ \alpha = 1/9, 1, $ and
$ 9 $.}
\end{figure}

Figure 2 shows the corresponding plots of $ \langle l \rangle $ versus 
$ \eps_r $.  Note that the localization lengths settle at the value of 
$ L_{rep}/2 $ at the transition, which is the expected finite $ L_{rep} $ 
behavior for the mutator phase.

\begin{figure}
\includegraphics[width = 0.6\linewidth, angle = -90]{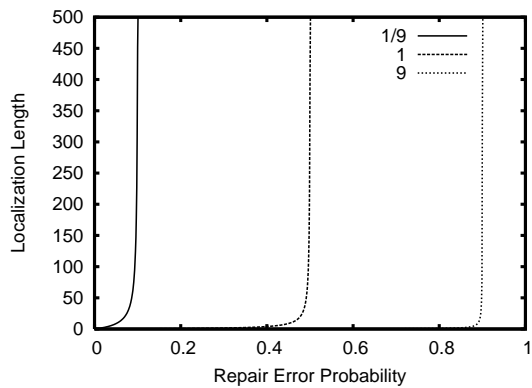}
\caption{Plots of $ \langle l \rangle $ versus $ \eps_r $ for $ \alpha = 1/9,
1, $ and $ 9 $.}
\end{figure}

Finally, we may use our model to estimate the equilibrium fraction of mismatch
repair deficient strains in {\it E. coli}.  The {\it E. coli} genome has 
$ 4,639,221 $ base pairs, comprising $ 4,403 $ genes \cite{ECOL1}.  Based on 
calculations for {\it Saccharomyces cerivisiae}, or Baker's yeast, we esimate 
that between $ 18 - 30\% $ of these genes are ``viability'' genes, i.e, 
required for {\it E. coli} survival \cite{YEAST1, YEAST2} (unfortunately, 
similarly detailed data is not currently available for {\it E. coli}, so we 
had to make an estimate based on available information).  Thus, we assume 
approximately $ 1,000 $ ``viability'' genes, which we gather into the 
viability peak of our model.  The mismatch repair system is controlled by the 
MutH, MutL, MutS, and UvrD (or MutU) proteins, giving $ 4 $ repair genes.  If 
we simply use the average gene length, and assume that the same fraction of 
base pairs must be conserved in both the viability genes and in the mismatch 
repair genes, then in our model we obtain $ \alpha \approx 1000/4 = 250 $.  
Since mismatch repair has a failure probability of $ 10^{-4} - 10^{-1} $ per 
mismatched base pair, we estimate an equilibrium fraction of mutators in the 
range of $ 4 \times 10^{-7} - 4 \times 10^{-4} $.  For {\it E. coli}, the 
observed equilibrium fraction of mutators is on the order of $ 10^{-5} - 
10^{-3} $ \cite{MUT2}.  While encouraging, our result is nevertheless based
on a number of simplifying assumptions.  The strongest evidence in support of 
our model would be the experimental observation of the repair catastrophe 
itself.  While it is not clear how to selectively control the efficiency of 
the mismatch repair system, if possible this would allow a direct experimental 
test of our model.

As a concluding remark, we should note that our prediction of a repair
catastrophe in mutator-nonmutator equilibrium suggests that phase
transitions may underlie the behavior of a variety of biological systems.
A classification of the phase behaviors inherent in various biological 
networks will greatly increase our understanding of the underlying dynamics 
governing such systems.

\begin{acknowledgments}

This research was supported by an NIH postdoctoral research fellowship,
and by the Howard-Hughes Medical Institute graduate research fellowship.
The authors would like to thank Dr. Andrew Murray and Ethan Perlstein 
for helpful conversations regarding this work.

\end{acknowledgments}

\bibliography{mu_catas}

\end{document}